\documentclass[preprint,aps,prd,floatfix,superscriptaddress]{revtex4-2}
\pdfoutput=1
\usepackage{graphicx}
\usepackage{subcaption}
\usepackage{mathrsfs}
\usepackage[colorlinks=true,
            urlcolor=blue,
            linkcolor=blue,
            citecolor=blue]{hyperref}
\usepackage[T1]{fontenc}
\begin{document} 

\title{Geometry of a generalized uncertainty-inspired spacetime}

\author{Douglas M. Gingrich}
\email{gingrich@ualberta.ca}
\affiliation{Department of Physics, University of Alberta, Edmonton,
  Alberta T6G 2E1, Canada}
\affiliation{TRIUMF, Vancouver, British Columbia V6T 2A3,
  Canada} 

\author{Saeed Rastgoo}
\email{srastgoo@ualberta.ca}
\affiliation{Department of Physics, University of Alberta, Edmonton,
  Alberta T6G 2E1, Canada}
\affiliation{Department of Mathematical and Statistical Sciences,
  University of Alberta, Edmonton, Alberta T6G 2G1, Canada}
\affiliation{Theoretical Physics Institute, University of Alberta,
  Edmonton, Alberta T6G 2E1, Canada} 

\date{\today}

\begin{abstract}
We examine the geometry of a generalized uncertainty-inspired quantum
black hole. 
The diagonal line element is not $t$-$r$ symmetric, i.e.\ $g_{00} \ne
-1/g_{11}$, which leads to an interesting approach to resolving the
classical curvature singularity. 
In this paper, we show, in Schwarzschild coordinates, the $r=0$
coordinate location is a null surface which is not a transition
surface or leads to a black bounce. 
We find the expansion of null geodesic congruences in the interior
turn around then vanishes at $r=0$, and the energy conditions are
predominately violated indicating a repulsive gravitational core.  
In addition, we show that the line element admits a wormhole solution
which is not traversable, and the black hole at its vanishing horizon
radius could be interpreted as a remnant. 
\end{abstract}
\maketitle
\section{Introduction}

Black holes solutions in general relativity (GR) are known to suffer
from singularities which are considered nonphysical.
To remove the curvature singularities modifications to GR are often
employed. 
If one considers the singularities as a breakdown of the classical 
description of gravity, the modifications are usually considered to
be quantum in origin.
Invoking generalized uncertainty principles (GUP) is one heuristic
approach to introducing quantum correction to the classical theory in
hope of resolving the singularity
issue~\cite{Kempf:1994su,Bosso:2023aht}. 
We investigate the spacetime of such an
approach~\cite{Fragomeno:2024tlh}.  

It is very difficult for modifications to GR to avoid having large
effects in low curvature regions.
If the modifications are quantum inspired, we can view the effects
beyond singularity resolution as effective quantum corrections.
As such, the quantum corrections should be small so as to not be
obviously observable.
The appropriate limits must reproduce the classical case.

It is not unlikely that increased knowledge of black holes could lead
to new physics.
Thus it is important to study effective quantum black holes to learn
more about the nature of quantum spacetime.
With the advent of observational data on gravitational mergers and 
imaging~\cite{Addazi:2021xuf,LISA:2022kgy,
LISACosmologyWorkingGroup:2022jok,AlvesBatista:2023wqm},
such programs of modified gravity are paramount.
The hope is that one day the data will be of sufficient quality to
test alternative theories of gravity and thus gain insight into the
nature of quantum spacetime.

It is common to maximally extend new black hole solutions and study
the properties of the spacetime~\cite{Modesto:2008im,
  Ashtekar:2020ckv, Liu:2021djf, Gambini:2020qhx, Alonso-Bardaji:2022ear}.
We start from the GUP-inspired spherically symmetric black hole metric
derived in~\cite{Fragomeno:2024tlh}. 
The black hole spacetime is asymptotically flat and the classical
singularity is resolved.
The theory admits two quantum parameters $Q_b$ and $Q_c$.
The parameter $Q_b$ introduces a distance scale $\sqrt{Q_b}$.
Assuming $Q_b$ is a small correction, it reduces the horizon radius by
a small amount. 
The parameter $Q_b$ is also responsible for causing the diagonal
line element to not be time-radius ($t$-$r$) symmetric, i.e.\ $g_{00}
\ne -1/g_{11}$.

The second quantum parameter, $Q_c$, introduces a mass dependent
distance scale $(Q_c m^2)^{1/8}$, where $m$ is the mass responsible
for generating the curvature of spacetime.
The parameter $Q_c$ affects all three metric components in the
spherically symmetric diagonal line element.
It causes two-spheres to have a minimum radius of $(Q_c m^2)^{1/8}$ and 
is responsible for resolving the classical singularity.
However, the $t$-$r$ asymmetry causes a coordinate singularity in the
diagonal line element at $r=0$.

One recognizes three distance, or mass, scales in the spacetime.
For the black hole solution, they are the black hole mass $m$, and two
effective quantum scales given in terms of the quantum parameters:
$\sqrt{Q_b}$ and $(Q_c m^2)^{1/8}$.
While $Q_b$ and $Q_c$ are unspecified by the theory, we expect them to
be small relative to $m$, else their observable effects would be
manifest.
For visualization and numerical work, we need to pick relative
numerical values for at least two of these three parameters.
We adapt the length scale hierarchy $(Q_c m^2)^{1/8} < \sqrt{Q_b} <
m$.
Starting with the standard choice of $m=1$, we pick $Q_b = 10^{-1} m^2$
which is about 16\% of the horizon radius, and $Q_c = 10^{-6} m^6$ which
is about 9\% of the horizon radius.
While these values, interpreted as quantum corrections, my be
considered large, none of our results will depend on these exact
values but they help make the visualizations manifest.
Throughout, we work in geometric units of $G = c = 1$.

The outline of this paper is as follows.
The background and formulas needed to analyze the spacetime are laid
out Sec.~\ref{sec:metric}. 
A sketch of the derivation leading to the GUP-modified metric
is given, along with more detailes in Appendix~\ref{appC}.
Radial geodesics and the scalar expansion of geodesic congruences
are developed, and the stress-energy tensor energy conditions are
stated. 
A derivation of the Painlev{\'e}-Gullstrand metric commonly used for
radial geodesics is given in Appendix~\ref{appA} for the
non-$t$-$r$-symmetric case.
The coordinate transformations needed to draw conformal diagrams are
given in Appendix~\ref{appB}. 
Different spacetimes corresponding to black hole, wormhole, and
remnant solutions are discussed in Sec.~\ref{sec:bh}, \ref{sec:wh},
and~\ref{sec:remnant}, respectively.
We summarize the findings in Sec.~\ref{sec:summary}.

\section{Generalized uncertainty spacetime\label{sec:metric}}

In this section, we outline the derivation leading to the GUP-modified
metric presented in~\cite{Fragomeno:2024tlh}. 
Starting from the black hole interior, solutions to the equations of
motions of the triad are found.
From these solutions, the interior metric is constructed.
The interior metric is then analytically extended to the full spacetime
by switching the timelike and radial spacelike coordinates.
The singularity is resolved, and checks made to ensure the correct
classical and asymptotic limits are obtained. 

The interior of a static spherically symmetric black hole expressed in
Ashtekar-Barbero variables using Schwarzschild coordinates is the
Kantowsk-Sachs~\cite{Ashtekar:2018cay} line element:

\begin{equation}
ds^2 = -N(\tilde{T})^2 d\tilde{T}^2 +
\frac{p_b^2(\tilde{t})}{L_0^2 |p_c(\tilde{t})|} d\tilde{r}^2 + 
|p_c(\tilde{t})| ( d\theta^2 + \sin^2\theta d\phi^2 )\, ,
\label{eq:metric}
\end{equation}
where $\tilde{t} \equiv \exp(\tilde{T})$ is timelike and $\tilde{r}$ is
spacelike in the interior.  
The components of the Ashtekar-Barbero connection and the densitized
triad are given by the configuration variables $b$ and $c$, and
associated conjugate momenta $p_b$ and $p_c$,
respectively~\cite{Ashtekar:2005qt}. 
Here $L_0$ is an infrared regulator.
None of the physical results depend on $L_0$ but on combinations with
other parameters which are all independent of the choice of $L_0$.
The lapse function $N$ is arbitrary, but a strategic choice is made
below. 

The algebra of the canonical variables, inherited from the algebra of 
the Ashtekar-Barbero connection and the densitized triad is

\begin{equation}
\{ b, p_b \} = \gamma \quad\textrm{and}\quad
\{ c, p_c \} = 2 \gamma\, ,
\label{eq:PoissonClass}
\end{equation}
where $\gamma$ is the Barbero-Immirzi parameter.
Modification to the Poisson algebra are made according to the GUP
approach by choosing a quadratic modification in the configuration
variables~\cite{Fragomeno:2024tlh}:

\begin{equation}
  \{ b, p_b \} = \gamma \left( 1 + \beta_b b^2 \right) \quad\textrm{and}\quad
  \{ c, p_c \} = 2 \gamma \left( 1 + \beta_c c^2 \right)\, ,  
\label{eq:PoissonGUP}
\end{equation}
where $\beta_b$ and $\beta_c$ are small dimensionless parameters,  
usually called GUP parameters.
The GUP correction is commonly written as an additional positive term in
the conjugate momentum squared.
Using the conjugate momentum in the GUP correction embodies a minimum
length scale. 
However, using the conjugate variable as in (\ref{eq:PoissonGUP}) embodies a
minimum momentum scale, which affects the triad and ultimately the
metric.
Since we are interested in GUP modifications of spacetime, we choose the latter.
We initially leave the sign of the GUP parameters unspecified.
However, negative signs must be chosen to maintain the signature and reality of
the metric over the entire domain $r \in (0,+\infty)$~\cite{Fragomeno:2024tlh}.
We acknowledge that this prescription no longer corresponds to a minimal
uncertainty principle but rather a GUP.  
As written in (\ref{eq:PoissonGUP}), the Poisson bracket does not lead to the
correct asymptotic limits or singularity resolution.
Guided by a prescription in loop quantum gravity~\cite{Ashtekar:2006wn,
Chiou:2008nm, Chiou:2012pg} in which the quantum parameters of the models are
made momentum dependent, we divide the GUP corrections by the conjugate momentum
squared~\cite{Fragomeno:2024tlh}:

\begin{equation}
  \{ b, p_b \} = \gamma \left( 1 + \frac{\beta_b b^2
    L_0^4}{p_b^2} \right) \quad\textrm{and}\quad
  \{ c, p_c \} = 2 \gamma \left( 1 + \frac{\beta_c c^2
    L_0^4}{p_c^2} \right)\, ,  
\label{eq:Poisson}
\end{equation}
where the dimensionality is absorbed into $L_0$. 
This represents a minimal modification that leads to consistent results
regarding classical and asymptotic limits, as well as singularity resolution. 
This deformed Poisson algebra is used to solve for the metric in
the black hole interior~\cite{Bosso:2020ztk,Fragomeno:2024tlh}
(see Appendix~\ref{appC}).

By choosing the lapse $N = \gamma\, \textrm{sgn}(p_c) \sqrt{|p_c|} /
b$, the classical equations of motion for $b$ and $p_b$ decouple from
those of $c$ and $p_c$.
Solving the classical Hamiltonian equations of motion, and replacing
the solutions for $p_b$ and $p_c$ in the metric (\ref{eq:metric})
yields the interior metric. 
The metric is analytically extended to the full spacetime by switching
$\tilde{t}\to r$ and $\tilde{r}\to t$. 

When studying the black hole metric, the following redefinitions are useful.

\begin{equation}
  Q_b = | \beta_b | \gamma^2 L_0^2 \quad\textrm{and}\quad
  Q_c = | \beta_c | \gamma^2 L_0^6\, .
  \label{eq:def}
\end{equation}

\subsection{Spacetime extension}

The interior of the Kruskal-Szekeres spacetime is isometric to the
Kantowski-Sachs vacuum solution (\ref{eq:metric})~\cite{Ashtekar:2018cay}.
For any choice of the time coordinate $\tilde{T}$ and the associated lapse
$\tilde{N}$, each point in the phase space defines a
metric~\cite{Ashtekar:2018lag}.
The metric has the natural coordinate range $-\infty < \tilde{T} <
\infty$ and $-\infty < \tilde{r} < \infty$. 

The black hole solution in the classical theory is valid in the region
$\tilde{T} > 0$.
In the quantum theory~\cite{Ashtekar:2018lag}, the singularity of the
classical theory is replace by a transition surface at $\tilde{T} =
0$, and $T < 0$ is a white hole interior.
The black hole interior solution can be analytically extended to
include $\tilde{T} = 0$, but we will not consider the white hole
spacetime. 

In the derivation of the metric~\cite{Fragomeno:2024tlh}, the variable
$\tilde{T}$ was transformed to the variable $\tilde{t}$ using the
transformation $\tilde{T} = \ln(\tilde{t})$ which naturally excludes
$\tilde{t} = 0$. 
In the exterior where $\tilde{t} \to r$ and $\tilde{r} \to t$, we see
$-\infty < t < \infty$, and $r = (0,\infty)$.

\subsection{Black hole metric}

The GUP-modified quantum black hole line element in Schwarzschild
coordinates is (see Appendix~\ref{appC} for a derivation)

\begin{equation}
  ds^2 = g_{00} dt^2 + g_{11} dr^2 + g_{22} d\Omega\, ,
  \label{eq:diagonal}
\end{equation}
where

\begin{eqnarray}
g_{00} & = & -\left( 1 + \frac{Q_b}{r^2} \right) \left( 1 + \frac{Q_c
m^2}{r^8} \right)^{-1/4} \left( 1 - \frac{2m}{\sqrt{r^2 +
Q_b}} \right)\nonumber\, ,\\ 
g_{11} & = & \left( 1 + \frac{Q_c m^2}{r^8} \right)^{1/4} \left( 1
- \frac{2m}{\sqrt{r^2 + Q_b}} \right)^{-1}\, ,\label{eq:functions}\\
g_{22} & = & r^2 \left( 1 + \frac{Q_c m^2}{r^8} \right)^{1/4}\nonumber\, ,
\end{eqnarray}
and $d\Omega^2 = d\theta^2 + \sin^2\theta d\phi$ is the standard
Riemannian metric on the unit radius two-sphere.
The mass of the black hole is $m$, and $Q_b$ and $Q_c$ are small real 
quantum parameters, where $Q_b$ has dimensions [L]$^2$ and $Q_c$
dimensions [L]$^6$.
The quantum parameters $Q_b$ and $Q_c$ are taken to be fixed in all
spacetime while $m$ is a constant of the motion.

\begin{figure}[htb]
\includegraphics[width=0.5\linewidth]{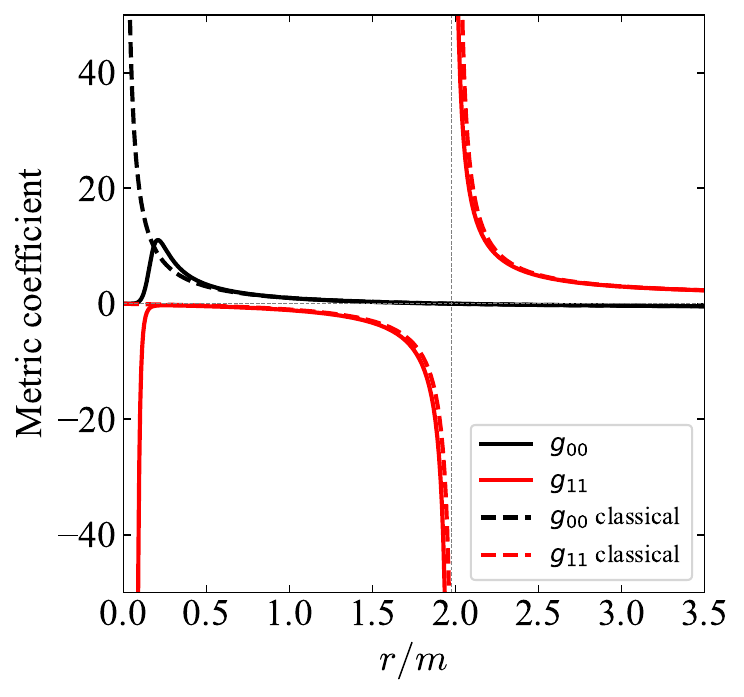}
\caption{\label{fig:metric}Metric components for the GUP-modified black
hole with $m = 1$, $Q_b = 0.1$, $Q_c = 10^{-6}$ (solid lines) and 
the classical black hole with $m = 1$ (dashed lines).}
\end{figure}

The coordinates have the natural domain $r \in (0,+\infty)$, $t\in
(-\infty,+\infty)$, $\theta \in [0,\pi]$, and $\phi \in [0,2\pi]$.
Figure~\ref{fig:metric} shows the metric components.
As $r\to \infty$, the metric is asymptotically flat in the same
sense as the Schwarzschild solution. 
In the classical limit, $Q_b, Q_c \to 0$; the solution is
Schwarzschild.
We observe that the classical singularity in $g_{00}$ is removed at
$r=0$, but a new coordinate singularity occurs in $g_{11}$ at $r=0$. 

As $r\to 0$, the component $g_{22}$ reaches a nonzero minimum $(Q_c
m^2)^{1/4}$, which is mass dependent.
We will show that under a certain condition between $m$ and $Q_b$, the
spacetime forms a wormhole with the throat of size $(Q_c m^2)^{1/4}$
located at $r = 0$. 
We could make a coordinate transformation to a new radial coordinate $r^2
\to (\bar{r}^8 - Q_c m^2)^{1/4}$ in which $\bar{r} > (Q_c m^2)^{1/8}$
but the other metric components become complex for $\bar{r} < (Q_b^4 -
Q_c m^2)^{1/8}$. 
We consider this frame an incomplete auxiliary frame, which is not the
physical one.



\subsection{Coordinate singularity}

We now examine the coordinate singularities in Schwarzschild
coordinates. 
Consider the hypersurfaces $r =$ constant.
The one-form normal to such hypersurfaces is $\partial r/\partial
x^\mu \equiv \partial_\mu r$.
The vector norm to the hypersurfaces is proportional to $g^{\mu\nu}
(\partial_\mu r)(\partial_\nu r) = g^{11}$.
We see that $g^{11}(r) = 0$ at $r = \sqrt{4m^2 - Q_b}$ and $r = 0$.
These are two null hypersurfaces.

For a spherically symmetric static spacetime, the Killing vector field
is $(\partial_t)^\mu$.
The norm is $g_{\mu\nu} (\partial_t)^\mu (\partial_t)^\nu = g_{00}$.
We see that $g_{00}(r) = 0$ at $r = \sqrt{4m^2 - Q_b}$.
Thus $r = r_h = \sqrt{4m^2 - Q_b}$ is an event horizon since it is
also a Killing horizon. 
The degenerate root (2nd order) occurring at $r=0$ is a coordinate
null singularity. 

\subsection{Different spherically symmetric static spacetimes}

The metric describes three different spherically symmetric static
geometries depending on the relative values of $m$ and $Q_b$. 
For $m > \sqrt{Q_b}/2$, a null surface occurs at $r_h =  \sqrt{(2m)^2 -
Q_b}$, which is a coordinate singularity and corresponds to an event
horizon. 
The event horizon location depends only on the single quantum variable
$Q_b$. 
This coordinate singularity does not occur for $m < \sqrt{Q_b}/2$ for 
any value of $r$. 
There is also the extremal case of $m = \sqrt{Q_b}/2$ with
characteristics depending on whether $m$ approaches $\sqrt{Q_b}/2$ from
above or below.
In addition, there is a null surface at $r = 0$ in all three cases.
All curvature invariants are finite over the full
spacetime~\cite{Fragomeno:2024tlh}.  
Thus the singularity at $r=0$ is a coordinate singularity which we
will study in the following sections.

In what follows, we will develop and calculate the radial
geodesics, scalar expansion of congruence of null geodesics,
stress-energy tensor, and energy conditions. 
These will then be used to separately study the black hole and
wormhole geometries.
\subsection{Geodesics}

To study the coordinate singularity at $r=0$, we will use radial
geodesics.
The geodesic equation with metric compatibility implies

\begin{equation}
\kappa =  -g_{\mu\nu} \frac{dx^\mu}{d\lambda}
\frac{dx^\nu}{d\lambda}\, ,
\end{equation}
where $\lambda$ is an affine parameter and $\kappa$ is a constant.
For massive particles $\kappa = 1$ and we set $\lambda = \tau$
the proper time.
For massless particles $\kappa = 0$ and $\lambda$ is not fixed.

For radial geodesics, $d\theta = d\phi = 0$ and the equation expands 
to

\begin{equation}
\frac{dr}{d\lambda} = \pm \left[ \frac{-g_{00}}{g_{11}}
\left( \frac{dt}{d\lambda} \right)^2 -
\frac{\kappa}{g_{11}}\right]^{1/2}\, .  
\end{equation}
Since the metric is static and spherically symmetric, there is one
asymptotically timelike Killing vector field $K^\mu = (\partial_t)^\mu
= (1,0,0,0)$ associated with energy

\begin{equation}
E = -K_\mu \frac{dx^\mu}{d\lambda} = -g_{00} \frac{dt}{d\lambda}\, ,
\end{equation}
where $E$ is a constant energy per unit mass.
For timelike radial geodesics,

\begin{equation}
\frac{dr}{d\tau} = \pm \left( \frac{E^2 + g_{00}}{-g_{00} g_{11}}
\right)^{1/2}
\quad\textrm{and}\quad
\frac{dr}{dt} = \pm \left[ -\frac{g_{00}}{g_{11}} \left( 
  \frac{E^2 + g_{00}}{E^2} \right) \right]^{1/2}\, .
\label{eq:radial}
\end{equation}
For massive particles starting at rest from infinity,
$E=1$. 
The $-g_{00}$ term in the numerator acts as a potential.

Alternatively, for null radial geodesics

\begin{equation}
\frac{dr}{d\lambda} = \pm \frac{E}{(-g_{00} g_{11})^{1/2}}
\quad\textrm{and}\quad
\frac{dr}{dt} = \pm \left( -\frac{g_{00}}{g_{11}} \right)^{1/2}\, .
\end{equation}
These radial geodesics have no effective potential and $E =
\hbar\omega$ is the energy of the massless particle.

For a radial infalling observer, one often uses the
Painlev{\'e}-Gullstrand form of the metric since $d\tau/dt = 1$.
For a non-$t$-$r$-symmetric diagonal metric one needs to be careful
not to have assumed $g_{00} = -1/g_{11}$.
Appendix~\ref{appA} presents a derivation of the
Painlev{\'e}-Gullstrand coordinates under no such assumption.
The resulting radial timelike geodesics are identical to
(\ref{eq:radial}). 
\subsection{Geodesic congruences}

We now examine the geodesic congruences of the GUP spacetime by
calculating the scalar expansion of a congruence of light rays in  
Kruskal-Szekeres coordinates (Appendix~\ref{appB} ) using null
geodesics. 
Here, ingoing refers to light rays moving on curves of constant
$V = V_0$, while outgoing designates light rays moving on
curves of constant $U = U_0$.
We note that if $V_0 > 0$, then $r$ decreases along the ingoing rays. 
While $r$ increases along the outgoing rays for $U_0 < 0$ in the
exterior and $U_0 > 0$ in the interior, and the horizon is located at
$U_0 = 0$. 
Outgoing and ingoing light rays have
\begin{equation}
k_\mu^+ = -\partial_\mu U \quad \textrm{and} \quad
k_\mu^- = -\partial_\mu V
\end{equation}
as their affinely parameterized tangent dual vectors, where $+$ refers
to outgoing and $-$ to ingoing.
The affine parameters are $\lambda_\pm = \mp r_*$.

For affinely parameterized tangent vectors, the scalar expansion can
be calculated using
\begin{equation}
\theta_\pm = (k^\mu_\pm)_{; \mu} = \frac{1}{\sqrt{-g}} \left(
\sqrt{-g}\,k_\pm^\mu \right)_{, \mu}\, ,
\end{equation}
where $g$ is the determinant of the Kruskal-Szekeres metric.
In Kruskal coordinates, only one component of the tangent vector is
nonzero,
\begin{equation}
k_+^\mu = (0,-1/|g_{UV}|,0,0) \quad \textrm{and} \quad
k_-^\mu = (-1/|g_{UV}|,0,0,0)\, .
\end{equation}
For these tangent vectors, the scalar expansions becomes
\begin{equation}
\theta_+ = -\frac{U_0}{4m} \frac{1}{\sqrt{-g_{00} g_{11}}}
\frac{g_{22}^\prime}{g_{22}} \quad \textrm{and} \quad 
\theta_- = -\frac{V_0}{4m} \frac{1}{\sqrt{-g_{00} g_{11}}}
\frac{g_{22}^\prime}{g_{22}}\, .
\end{equation}
The $g_{\mu\nu}$ metric coefficients are in Schwarzschild coordinates
and the prime denotes differentiation with respect to $r$.
In the classical limit, $\sqrt{-g_{00} g_{11}} = 1$ and
$g_{22}^\prime/g_{22} = 2/r$.
The ingoing expansion scalar for the GUP metric coefficients is
\begin{equation}
\theta_- = -\frac{V_0}{2mr} \left( 1+\frac{Q_b}{r^2} \right)
^{-1/2}\left( 1 + \frac{m^2 Q_c}{r^8}\right)^{-1}\, . 
\end{equation}
The corresponding outgoing expansion scalar is obtained by replacing
$V_0$ with $U_0$.
We are reminded that while $V_0$ is always positive, for outgoing
rays, $U_0$ is negative outside the black hole and positive inside the
black hole. 

The rate of change in the scalar expansions are
\begin{equation}
\frac{d\theta_+}{d\lambda} = -\frac{U_0}{k} \left(
\frac{1}{\sqrt{-g_{00} g_{11}}} \frac{g_{22}^\prime}{g_{22}}
\right)^\prime \sqrt{\frac{-g_{00}}{g_{11}}}
\quad \textrm{and} \quad  
\frac{d\theta_-}{d\lambda} = \frac{V_0}{k} \left(
\frac{1}{\sqrt{-g_{00} g_{11}}} \frac{g_{22}^\prime}{g_{22}}
\right)^\prime \sqrt{\frac{-g_{00}}{g_{11}}}\, , 
\end{equation}
The ingoing expansion scalar for the GUP metric coefficients is
\begin{eqnarray}
\frac{d\theta_-}{d\lambda} & = & -\frac{V_0}{2mr^2}
\left( \sqrt{1 + \frac{Q_b}{r^2}} - \frac{2m}{r} \right)
\left[ 1 -\frac{m^2 Q_c}{r^8} \left( 7 + \frac{8Q_b}{r^2} \right)
  \right]\nonumber\\
& & \times \left( 1 + \frac{Q_b}{r^2} \right)^{-3/2}
\left( 1 + \frac{m^2 Q_c}{r^8} \right)^{-9/4}\, ,
\end{eqnarray}
and the corresponding outgoing expansion scalar rate of change is
obtained by replacing $V_0$ with $U_0$. 
\subsection{Stress-energy tensor energy conditions}

The effective equations of motion are not the Einstein equations.
However, the Einstein tensor of the effective quantum metric can provide
useful information about the energy conditions and suitable
interpretations related to the singularity
resolution~\cite{Curiel:2014zba}. 
In the absence of an underlying theory, we take the solution governed
by the Einstein equations equipped with an effective energy-momentum
tensor.  
That is, the quantum-corrected metric in vacuum is applied to the
left-hand side of the Einstein equations to give a nonzero effective 
energy-momentum tensor on the right-hand side of the equations.
There is no physical matter field with the effective stress-energy of
our vacuum solutions.
This effective matter field is expected to violate the energy
conditions. 

Energy conditions have physical, geometric, and effective
formulations~\cite{Curiel:2014zba}.
The geometric formalism is in terms of the geometric tensors and the 
physical formalism in terms of the stress energy tensor itself.
These interpretations are equivalent to each other in any theory that
can be formulated with effective Einstein equations.
For the reasons stated above, the physical meanings have no validity
in this case.
The concept of the effective stress-energy tensor is useful because
its geometrical relationship with the spacetime curvature.
Here, we only consider the operational definitions. 

The stress energy tensor is symmetric, and in this case, diagonal and
is in the type-I canonical form~\cite{HawkingEllis}.
The eigenvalues can be viewed as an energy density and three 
principle pressures, assume the quantum corrections to Einstein
gravity are minimally coupled to an anisotropic perfect fluid
form~\cite{Cho:2017nhx}. 
Effectively, it is the anisotropic fluid that drives the quantum
corrections.
Because of the symmetries of the background spacetime, the effective
energy-momentum tensor of this anisotropic perfect fluid can be
written as 

\begin{equation}
T_{\mu\nu} = (\rho + p_2) u_\mu u_\nu + (p_1-p_2) x_\mu x_\nu + p_2
g_{\mu\nu}\, ,
\end{equation}
where $\rho$ is the energy density measured by a comoving observer
with the fluid, and $p_1$ and $p_2$ are the radial and tangential
pressures, respectively.
Here $u^\mu$ is the timelike four-velocity, $x^\mu$ is the
spacelike unit vector orthogonal to $u^\mu$ and the angular
directions, and $g_{\mu\nu}$ is the metric of the background
spacetime. 
That is

\begin{equation}
  u_\mu u^\mu = -1\, , \quad
  x_\mu x^\mu = 1\, , \quad \textrm{and} \quad
  u_\mu x^\mu = 0\, .
\end{equation}
The nonzero components from the Einstein equations for the
perfect fluid yield (in the exterior)

\begin{equation}
  T_0^{\ 0} = -\rho\, , \quad
  T_1^{\ 1} = p_1\, , \quad \textrm{and} \quad
  T_2^{\ 2} = T_3^{\ 3} = p_2 = p_3\, .
\label{eq:pressures}
\end{equation}
The later equality is a consequence of spherical symmetry.
Later, we will see that as $r\to\infty$ all $T_\mu^{\ \mu}\to 0$,
recovering the asymptotic limit.

The effective energy conditions are given by

\bigskip
\noindent
\begin{tabular}{lllll}
\textrm{null energy condition}     & \textrm{(NEC)} &
$\rho + p_i \ge 0$\, , & &\nonumber\\
\textrm{weak energy condition}     & \textrm{(WEC)} &
$\rho \ge 0$ & \textrm{and} & $\rho + p_i \ge 0$\, ,\nonumber\\
\textrm{strong energy condition}   & \textrm{(SEC)} &
$\rho + p_i \ge 0$ & \textrm{and} & $\rho + \sum_i p_i \ge 0$\, ,\nonumber\\
\textrm{dominant energy condition} & \textrm{(DEC)} &
$\rho \ge |p_i|$\, , & &\nonumber
\end{tabular}
\bigskip

\noindent
for each $i=1,2,3$.
These are sometimes referred to as the point energy conditions.
We will not consider averaged energy conditions~\cite{Visser} here.
Demonstrating that the NEC is violated is sufficient to conclude that
the WEC, SEC, and DEC  will also be violated.

The expressions in (\ref{eq:pressures}) are for the exterior region.
In the interior the $t$ and $r$ coordinates swap their
timelike/spacelike characters.
For the interior region we switch $T_0^{\ 0} \leftrightarrow
-T_1^{\ 1}$.  
Hence the reader does not need to modify $\rho$ and $p_i$ in our plots
as they cover all $r$ for black holes.

\section{Black hole geometry\label{sec:bh}}

\begin{figure}[htb]
\begin{subfigure}[c]{0.44\linewidth}
\includegraphics[width=\linewidth]{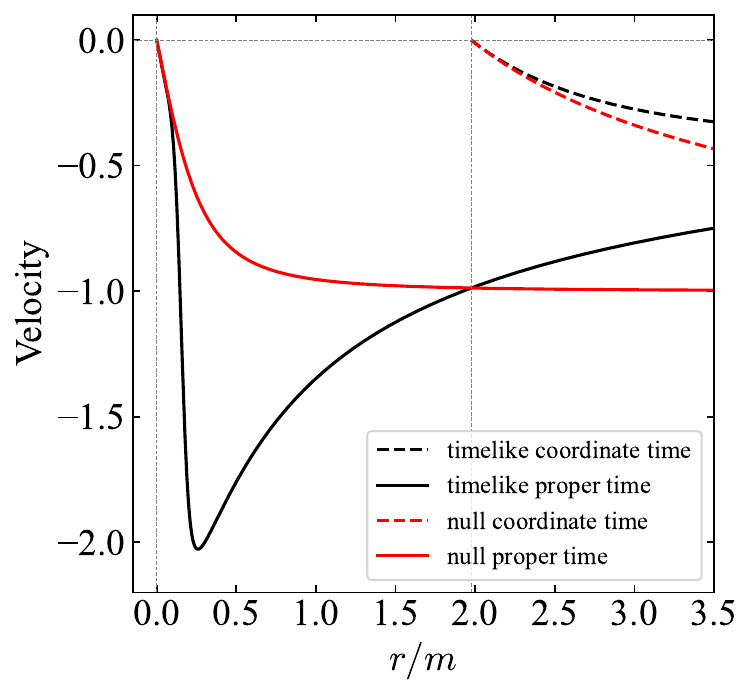}
\caption{\label{fig:geo_black}Radial geodesics with $E = 1$.}
\end{subfigure}
\hfill
\begin{subfigure}[c]{0.49\linewidth}
\includegraphics[width=\linewidth]{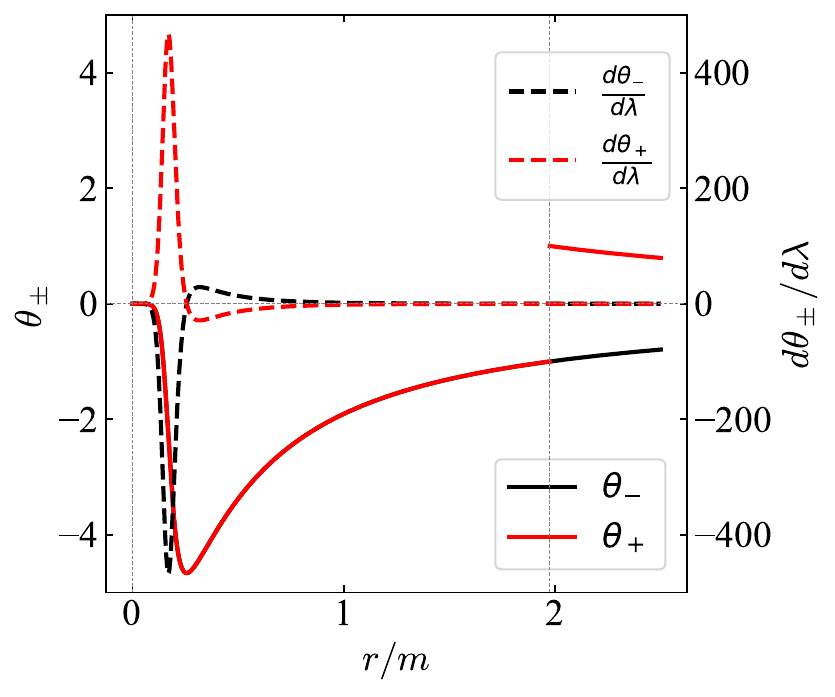}
\caption{\label{fig:expan_black}Expansion scalars.}
\end{subfigure}
\caption{Radial geodesics and expansion scalars for $m > \sqrt{Q_b}/2$ 
with $m = 1, Q_b = 0.1$, and $Q_c = 10^{-6}$.
For the expansion scalars $V_0 = |U_0| = 4m$, and the solid black line
lies under the solid red line in the black hole interior.} 
\end{figure}

The black hole solution has two coordinate singularities: $r = r_h$
and $r = 0$.
All scalar curvature invariants are finite at these
values~\cite{Fragomeno:2024tlh}. 
Figure~\ref{fig:geo_black} shows radial geodesics' velocity for the
black hole spacetime.
We see that the timelike and null velocities both vanishes at the
horizon and at $r = 0$ in Schwarzschild coordinates.   
As in the classical case, massive and massless free particles appear
to stop at the event horizon when viewed by an external observer, but
otherwise pass through the horizon as viewed by the particle itself.
For the metric studied here, both massive and massless free
particles take an infinite time to reach the origin at $r = 0$ in
their own frame. 
We associate $r = 0$ with future null/timelike infinity; $r=0$ behaves
both like $\mathscr{I}^+$ and $i^+$, respectively. 
The geodesics never reach the $r=0$ surface and avoid the coordinate
singularity. 
The change in the afffine parameter $\Delta\lambda$ never vanishes and
the geodesics are complete.

Figure~\ref{fig:expan_black} shows the expansion scalars and their
rate of change for ingoing and outgoing null geodesics.
The ingoing expansion $\theta_-$ is always negative while $\theta_+$
chances sign on opposite sides of the horizon.
The horizon radius is thus a trapped surface.
The asymptotic behavior is $\theta_\pm\to 0$ and
$d\theta_\pm/d\lambda\to 0$ as $r\to\infty$ which is identical to the
classical case.  
For the GUP black hole, $\theta_\pm\to 0$ and $d\theta_\pm/d\lambda\to
0$ as $r\to 0$.
This behaviour is identical to that at future null infinity and there
is no caustic of the congurances.
More importantly, in the interior $\theta_\pm$ decrease as the radial
coordinate decreases, reaching a minimum value at a turning point
in the interior after which gravity becomes effectively repulsive at
small $r$ and stays so until $r = 0$. 
Qualitative similar behaviour is obtained for timelike geodesics and
null geodesics in Painlev{\'e}-Gullstrand
coordinates~\cite{Rastgoo:2022mks,Fragomeno:2024tlh}.   

\begin{figure}[htb]
\begin{subfigure}[b]{0.49\linewidth}
\includegraphics[width=\linewidth]{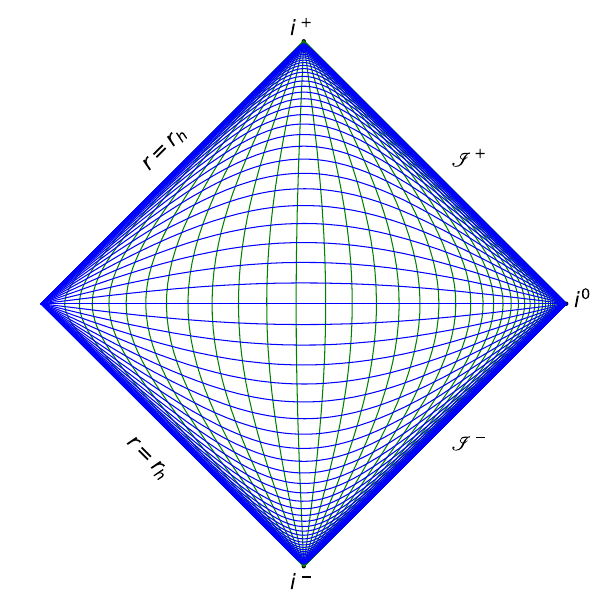}
\caption{\label{fig:pen_external}Exterior.}
\end{subfigure}
\hfill
\begin{subfigure}[b]{0.49\linewidth}
\includegraphics[width=\linewidth]{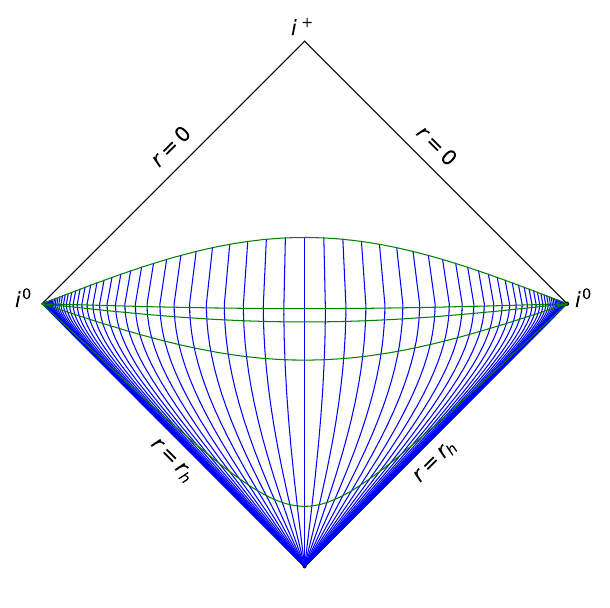}
\caption{\label{fig:pen_internal}Interior.}
\end{subfigure}
\caption{\label{fig:penrose}Carter-Penrose diagrams (not maximally extended).
The geodesics are drawn equidistant in the coordinates. 
The bifurcation sphere is not labeled, and the interior patch should be
glued to the exterior patch along the horizon between the bifurcation
point and $i^+/i^0$.}
\end{figure}

Using the results of Appendix~\ref{appB}, the Carter-Penrose
diagrams are shown in Fig.~\ref{fig:penrose}.  
The Carter-Penrose diagram for the exterior patch
(Fig.~\ref{fig:pen_external}) is similar to the Schwarzschild case.
The timelike geodesics (green) have been drawn from $r = r_h$ (within
the numerical resolution) increasing in constant $r$ in steps of 0.5.
The spacelike geodesics (blue) have been drawn for $t=0$ increasing
and decreasing in constant $t$ in steps of 0.5.
The bifurcation point (sphere) on the left is not labeled.
For large $r$, $r_* \approx r$, and the timelike geodesics appear
approximately equally spaced.
The Carter-Penrose diagram for the interior patch
(Fig.~\ref{fig:pen_internal}) is unique to this metric.
The timelike geodesics (blue) have been drawn for $t=0$ increasing and
decreasing in constant $t$ in steps of 0.5.
The spacelike geodesics (green) have been drawn between $r > 0$ and $r_h$
in steps of 0.5.
The geodesics never reach the $r = 0$ null surface due to the finite
$Q_b = 0.1$ value.
For intermediate $r$, $r_* \approx$ constant and the spacelike
geodesics appear closely spaced.
The spacelike geodesics are widely spaced near $r_h$ and $r=0$ where
$r_*$ is changing rapidly.
  
In the interior, the horizontal line (spacelike surface) at the
top of the classical conformal diagram that correspond to the
singularity at $r=0$ has been replaced by future null infinity
$\mathcal{J}^+$ and future timelike infinity $i^+$.
The two Kruskal-Szekeres spacetimes~\footnote{The Kruskal-Szekeres
maximally extended Schwarzschild solution is commonly divided into
four regions: I exterior, II interior black hole, III parallel
exterior, and IV interior white hole. We are referring to the regions
I/II and III/IV as the two spacetimes.} are causally disconnected as
it is not possible to move between them in a finite amount of time.    

\begin{figure}[htb]
\includegraphics[width=\linewidth]{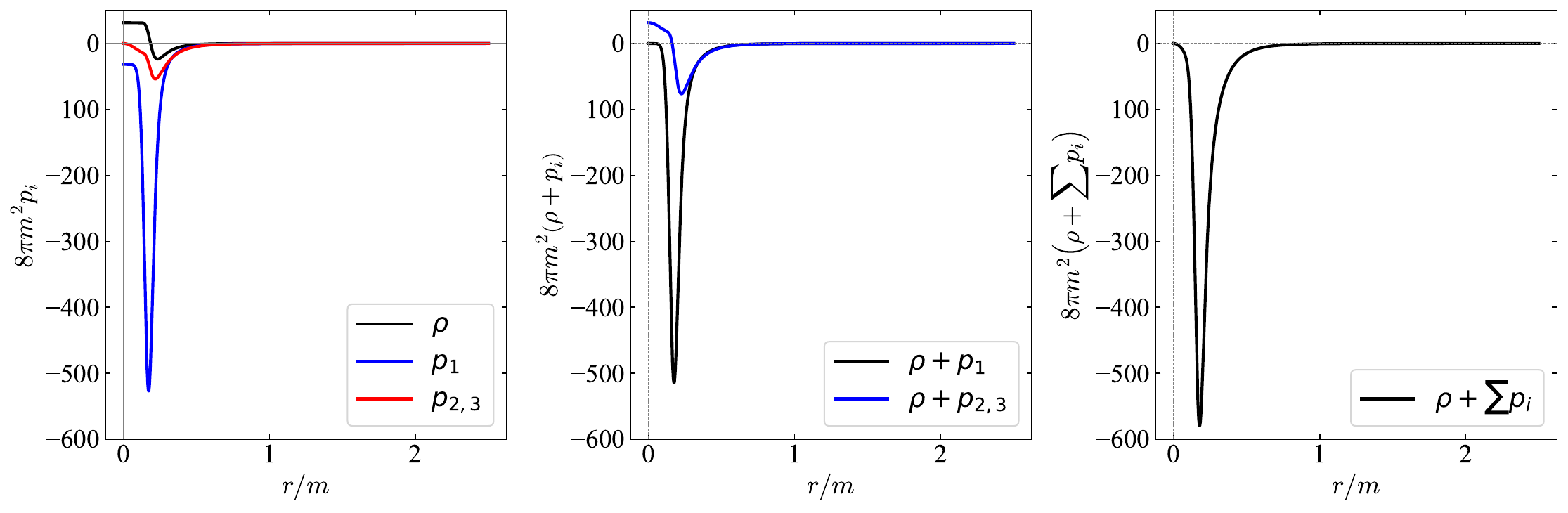}
\caption{\label{fig:energy_black}Density and pressure combinations
  needed to determine the energy conditions for $m >
  \sqrt{Q_b}/2$ with $m = 1, Q_b = 0.1$, and $Q_c = 10^{-6}$.}
\end{figure}

Figure~\ref{fig:energy_black} shows the quantities necessary to
evaluate the energy conditions.
Since $|p_1| > |\rho|$, $\rho+p_1 < 0$, and $\rho+\sum p_i < 0$ occur
for some finite value of $r$, all the energy conditions are violated. 
On the other hand, all the energy conditions are satisfied at $r=0$
and $r\to\infty$. 
Although not clearly visible in Fig.~\ref{fig:energy_black}, all the
energy conditions are also obeyed at the horizon.

\section{Wormhole geometry\label{sec:wh}}

The line element (\ref{eq:diagonal}) and (\ref{eq:functions}) looks
similar to the Simpson-Visser line element~\cite{Simpson:2018tsi}
suggesting the investigation of a possible wormhole.
Consider the case when $m < \sqrt{Q_b}/2$.
For all $r$, the factor $(1 - 2m/\sqrt{r^2+Q_b})$ in the metric
functions (\ref{eq:functions}) will be positive nonzero.
Therefore, no event horizon exist.
The spacetime is asymptotically flat for $r\to\pm\infty$ and only
$g_{11}$ admits a quadratic coordinate singularity at $r=0$.

When studying the timelike radial geodesics, we observe that
$g_{00}$ provides an effective potential due to the quantum parameter
$Q_b$ and $Q_c$ that would not be present in the Simpson-Visser case. 
The particle's energy squared $E^2$ needs to be high enough to get
over this potential barrier.
This happens for both coordinate and proper velocities for a massive
particle.
We interpret this condition as the energy needed to reach the wormhole
throat due to the quantum effect from $Q_b$ and $Q_c$. 

\begin{figure}[htb]
\begin{subfigure}[c]{0.43\linewidth}
\includegraphics[width=\linewidth]{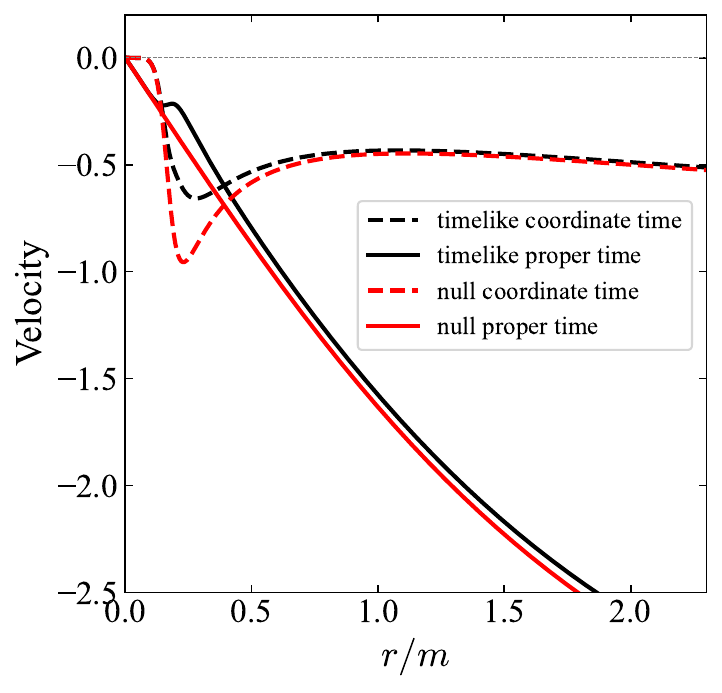}
\caption{\label{fig:geo_worm}Radial geodesics with $E=4$.} 
\end{subfigure}
\hfill
\begin{subfigure}[c]{0.49\linewidth}
\includegraphics[width=\linewidth]{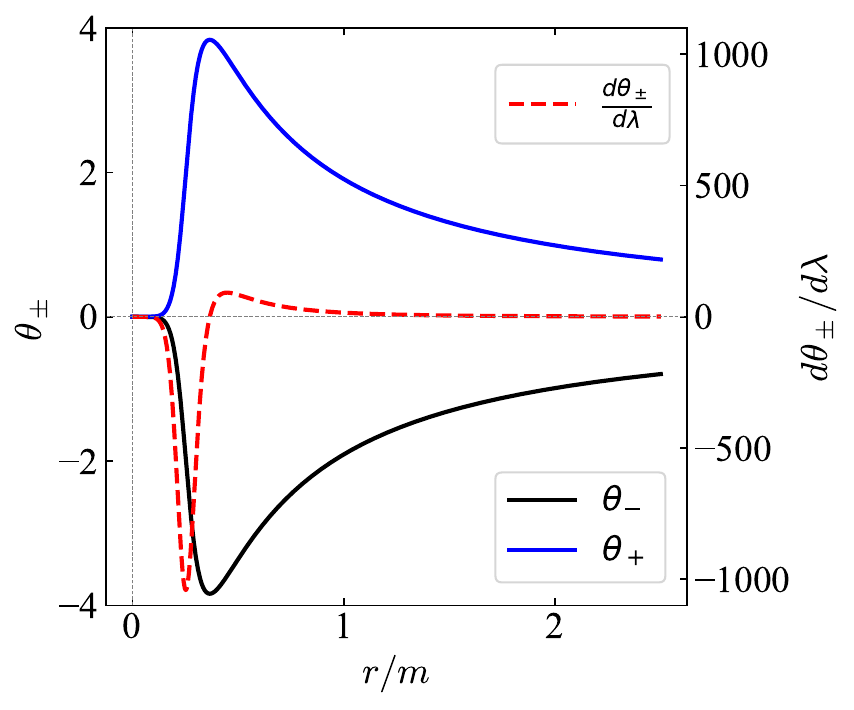}
\caption{\label{fig:expan_worm}Expansion scalars.} 
\end{subfigure}
\caption{Radial geodesics and expansion scalars for $m < \sqrt{Q_b}/2$ 
with $m = 1, Q_b = 5$, and $Q_c = 10^{-6}$.
For the expansion scalars $V_0 = |U_0| = 4m$.}
\end{figure}

Assuming the particle has enough energy to get over the effective
potential barrier, the radial geodesics are shown in
Fig.~\ref{fig:geo_worm}. 
The time for massive and massless free particles to reach the $r=0$
surface measured in its own frame is infinite. 
This is unlike an event horizon and we associate $r=0$ with
timelike/null infinity. 
The surface $r=0$ is not an event horizon and the geometry is not a 
degenerate black hole; nor is the geometry a traversable wormhole
(two-way or one-way). 
At best, we can refer to the spacetime as a nontraversable wormhole
that joins two casually disconnected spacetimes.
Otherwise, the wormhole is smoothly connected at the throat located at
$r=0$, with radius $(Q_c m^2)^{1/8}$.
The throat radius depends only on the quantum parameter $Q_c$ and is
mass dependent.
We are reminded that the size of the wormhole throat is on the order
of the Planck length for small masses.

Figure~\ref{fig:expan_worm} shows the expansion scalars and their
rate of change for ingoing and outgoing null geodesics.
The ingoing expansion $\theta_-$ is always negative while $\theta_+$
is always poitive.
The asymptotic behavior is $\theta_\pm\to 0$ and
$d\theta_\pm/d\lambda\to 0$ as $r\to\infty$ which is identical to the
classical case.  
For the GUP wormhole, $\theta_\pm\to 0$ and $d\theta_\pm/d\lambda\to
0$ as $r\to 0$.
This behaviour is identical to that at future null infinity and there
is no caustic of the congurancies.
More importantly, the absolute value of $\theta_\pm$ increases as the
radial coordinate decreases, reaching a minimum value at the a turning
point after which gravity becomes repulsive at small $r$ and 
stays so until $r = 0$. 

\begin{figure}[htb]
\begin{subfigure}[c]{0.43\linewidth}
\includegraphics[width=\linewidth]{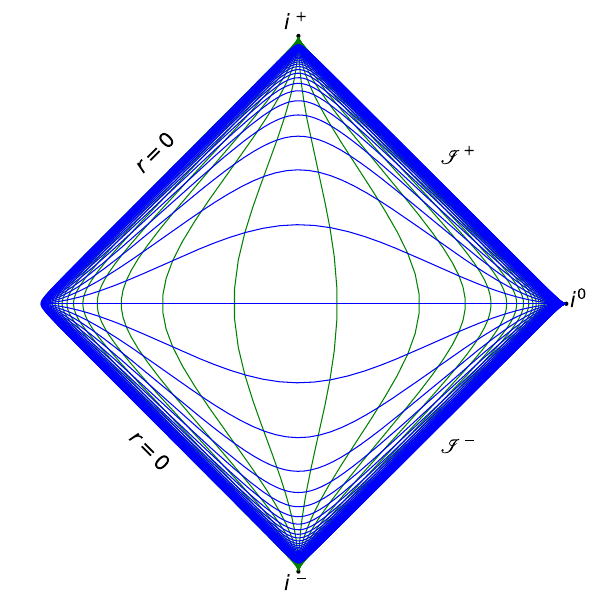}
\caption{\label{fig:pen_worm}Carter-Penrose diagram.} 
\end{subfigure}
\hfill
\begin{subfigure}[c]{0.42\linewidth}
\includegraphics[width=\linewidth]{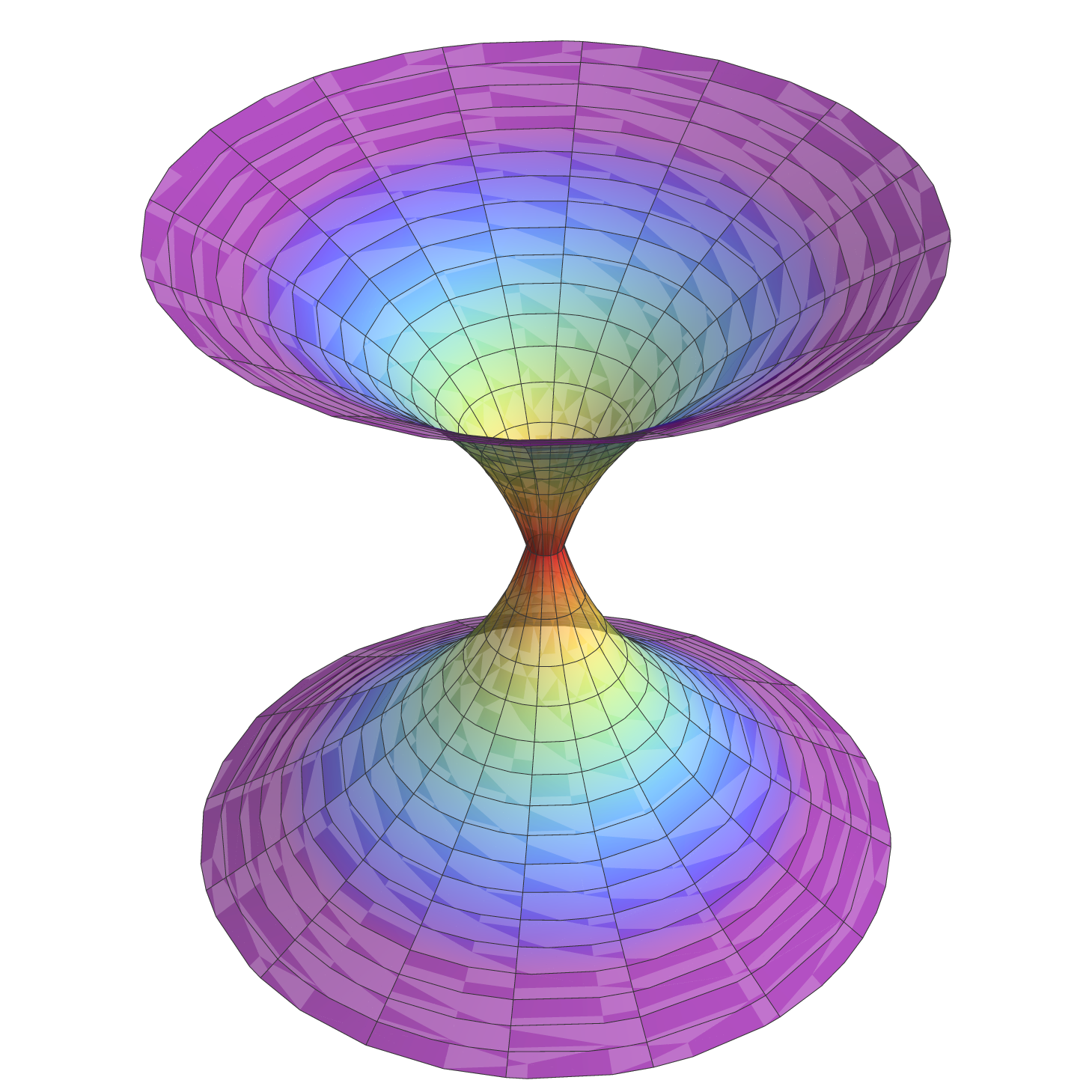}
\caption{\label{fig:embed}GUP wormhole with $m=1$, $Q_b = 5$, and
$Q_c = 10^{-6}$.} 
\end{subfigure}
\caption{Spacetime for $m < \sqrt{Q_b}/2$.}
\end{figure}

Figure~\ref{fig:pen_worm} shows a Carter-Penrose diagram for the
case of the wormhole. 
Figure~\ref{fig:embed} shows a embedding diagram for the
case of the wormhole.
The wormhole throat is not traversable in reality, or in practice. 

\begin{figure}[htb]
\includegraphics[width=\linewidth]{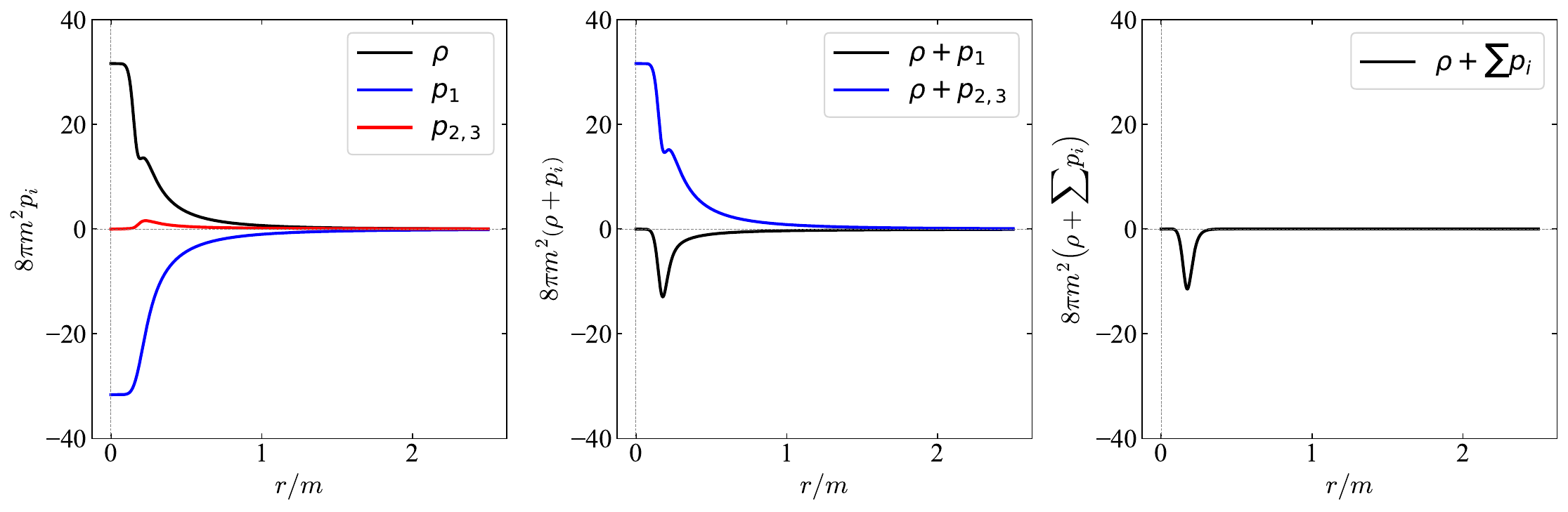}
\caption{\label{fig:energy_worm}Density and pressure combinations
needed to determine the energy conditions for $m < \sqrt{Q_b}/2$ with
$m = 1, Q_b = 5$, and $Q_c = 10^{-6}$.}  
\end{figure}

Figure~\ref{fig:energy_worm} shows the quantities necessary to
evaluate the energy conditions.
Since $|p_1| > \rho$, $\rho+p_1 < 0$, and $\rho+\sum p_i < 0$ occur
for some finite value of $r$, all the energy conditions are violated. 
On the other hand, all the energy conditions are satisfied at $r=0$
and $r\to\infty$.

It is of intrinsic interest to determine the mass at which the size of
the horizon radius $\sqrt{4m^2-Q_b}$ becomes the same as the minimum
size of the two-sphere $(Q_c m^2)^{1/8}$ in Schwarzschild coordinates.
This interesting surface can be viewed as the one in which the horizon
is hidden behind the wormhole throat~\cite{Modesto:2008im}.
The equation satisfying this equality is

\begin{equation}
16 m^4 - 8 Q_b m^2 - Q_c^{1/2} m + Q_b^2 = 0\, .
\label{eq:wormhole}
\end{equation}
Two solutions for $m$ are complex and one is negative.
The interesting real-positive solution is an expression in $Q_b$ and
$Q_c$ which is not simplifiable. 
This is because $Q_b$ and $Q_c$ are dimensional parameters of length
to the power of two and six respectively.  
To simplify the expression for $m$, we rewrite the dimension parameters
as $Q_b = \gamma^2 \beta_b L_0^2$ and $Q_c = \gamma^2 \beta_c L_0^6$,
which allows the common length scale to be factored out, leaving us
with ratios of dimensionless parameters. 
If we take the limit $\beta_c\to 0$, while $\beta_b$ is finite, we are
only able to satisfy the condition of (\ref{eq:wormhole}) at the
remnant mass give by $m = \sqrt{Q_b}/2$. 
However, for $\beta_c$ finite and $\beta_b\to 0$, we obtain

\begin{equation}
m \approx \frac{1}{2} \left( \frac{Q_c}{4} \right)^{1/6}\, .
\end{equation}
For this to occur before the black hole remnant mass, we need $Q_c
> 4 Q_b^3$. 

\section{Black hole remnant\label{sec:remnant}}

If the geometry starts as a black hole solution, the black hole may 
Hawking evaporate causing the black hole mass to decrease.
In the quantum regime when $m$ reaches $\sqrt{Q_b}/2$, the event
horizon vanishes and the black hole ceases to exist and a remnant is
form. 
The fate of the remnant can be determine by black hole thermodynamics.

Since the energy conditions are obeyed at the horizon, we might expect
the surface gravity to be well defined.
We will now calculate the surface gravity for a non-$t$-$r$-symmetric
black hole. 
The surface gravity can be written as

\begin{equation}
\kappa = V a = \sqrt{\nabla_\mu V \nabla^\mu V}
\end{equation}
evaluated at the horizon.
In this expression, $V$ is the magnitude of the asymptotically
timelike Killing vector field or the redshift factor, and $a$ is the
magnitude of the four-acceleration. 
The Killing vector and four-velocity of a static observer are

\begin{equation}
K^\mu = (1,0,0,0) \quad \textrm{and} \quad
u^\mu = \left( \sqrt{-g_{00}},0,0,0 \right)\, . 
\end{equation}
The four-acceleration and its magnitude are

\begin{equation}
a_\mu = \frac{1}{2g_{00}} \frac{d g_{00}}{dr} \delta^r_\mu
\quad \textrm{and} \quad 
a = \frac{1}{2\sqrt{g_{11}} g_{00}} \frac{d g_{00}}{dr}\, .
\end{equation}
Thus the surface gravity is

\begin{equation}
\kappa = \left. \sqrt{-g_{00}}\ a\right|_{r=r_h} = \left.
\frac{-1}{2\sqrt{-g_{00} g_{11}}} \frac{d g_{00}}{dr} \right|_{r=r_h}\, . 
\end{equation}

As a check, a second way to calculate the surface gravity is to use
the redshift factor $V = \sqrt{-g_{00}}$ alone.
The surface gravity is 

\begin{equation}
\kappa = \sqrt{\nabla_\mu
  V\nabla_\nu V g^{\nu\mu}} = \partial_r V \sqrt{g^{11}} =
\frac{-1}{2\sqrt{-g_{00} g_{11}}} \frac{d g_{00}}{dr}\, .
\end{equation}
Assuming the Hawking temperature maintains its meaning in the quantum
regime, the Hawking temperature is identified with

\begin{equation}
T = \frac{\kappa}{2\pi} =  \frac{1}{8\pi m}  
\left( 1 + \frac{Q_c m^2}{r_h^8} \right)^{-1/4}\, .
\end{equation}
The temperature depends on the mass as in the classical case but the
quantity in brackets modifies the usual Schwarzschild temperature.
The temperature is thus corrected by both quantum parameters.
Typically there is a $1/r_h$ dependence for the black hole under
consideration in front of the bracket.
Here, this is cancelled by the extra factor 
in the $g_{00}$ metric component.

\begin{figure}[htb]
\begin{subfigure}[c]{0.49\linewidth}
\includegraphics[width=\linewidth]{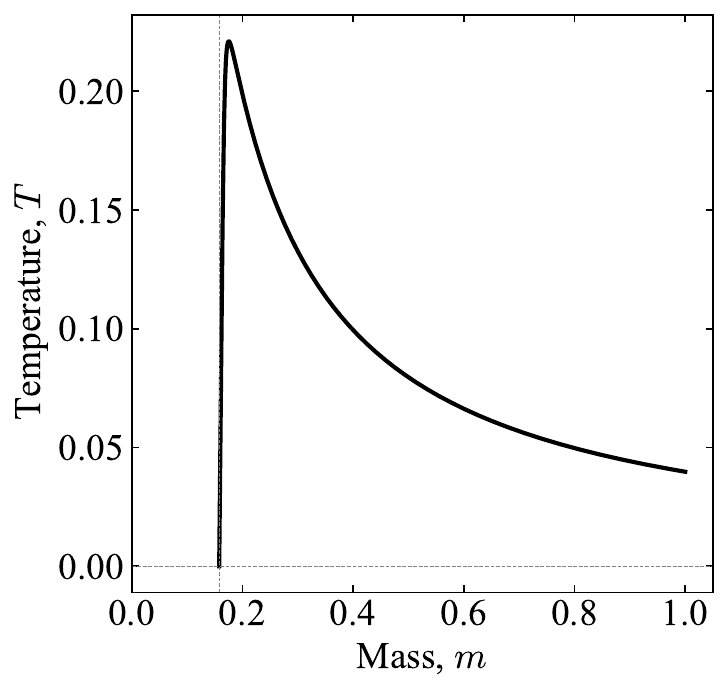}
\caption{\label{fig:temperature}Temperature, $T$.}
\end{subfigure}
\hfill
\begin{subfigure}[c]{0.49\linewidth}
\includegraphics[width=\linewidth]{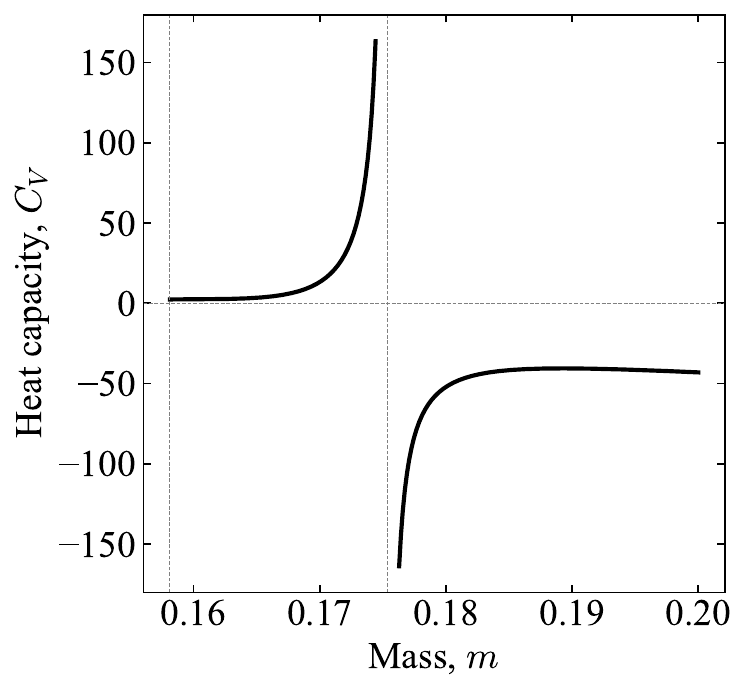}
\caption{\label{fig:heat}Heat capacity, $C_V$.}
\end{subfigure}
\caption{Thermal properties of the black hole for $Q_b = 0.1$ and $Q_c =
  10^{-6}$.}   
\end{figure}

This temperature is plotted in (\ref{fig:temperature}).
We observe that in contrast to the Schwarzschild case, there exist a
maximum temperature and that the temperature vanishes at the minimum
mass $\sqrt{Q_b}/2$.
The same result could have been obtained in Euclidean space by
applying a Wick rotation of the metric in the exterior
region~\cite{Ashtekar:2023cod}. 

To study the thermodynamics further, we calculate the heat capacity
using

\begin{equation}
C_V = \left( \frac{\partial T}{\partial m} \right)^{-1}\, .
\end{equation}
The heat capacity versus mass is plotted in (\ref{fig:heat}).
For large masses, we observe that $C_V$ is negative and asymptotes to
the classical value.
At the mass corresponding to the maximum temperature there is a phase
transition and $C_V$ changes sign.
For black holes with a mass less than the maximum-temperature mass, we
find that $C_V > 0$ and the system is thermodynamically stable.
The stability statement is based purely on the classical arguments of
a minimum mass (energy), zero temperature, and positive heat capacity.  

Further considerations are necessary if the quantum parameter
$\sqrt{Q_b}$ is close to the Planck scale.
Classically, the black hole will Hawking evaporate down to the Planck
length at which point it will become thermodynamically stable with
finite mass the order of the Planck mass. 
This is an ideal situation which ignores many effects and unknowns. 
For example, the black hole could come to thermal equilibrium with
the cosmic microwave background before its temperature vanishes.
In addition, we have not studied the black hole decay time as this
would require greybody factors and the standard decay time
calculations may not even be applicable in the deep quantum regime.
Near the Planck scale the decay time may well approach infinity.
Furthermore, in the final stage of evaporation when the horizon
radius reaches the Planck length, the spacetime fluctuations of the
manifold probably become comparable in length.
It is unknown if quantum effects will destabilize the remnant.
All we can say is that thermodynamic stability is necessary, but not
sufficient, to claim a stable remnant.

It is interesting to calculate the entropy and compare it with the
classical expression.
The entropy can be obtained from

\begin{equation}
  S = \int \frac{dm}{T} = 8\pi \int m \left( 1 +
  \frac{Q_c m^2}{r_h^8} \right)^{1/4} dm\, .
\label{eq:entropy}
\end{equation}
We can also calculate the area of the black hole at the horizon by
setting $r=r_h$ and $dt = dr = 0$: 

\begin{equation}
A = \left. \int g_{22} \sin\theta d\theta d\phi \right|_{r=r_h} =
4 \pi r_h^2 \left( 1 + \frac{Q_c m^2}{r_h^8} \right)^{1/4}\, .
\end{equation}
In both expressions $r_h$ is an implicit function of $m$.
Unlike the horizon radius, which depends only on the quantum parameter
$Q_b$, the entropy (also temperature and heat capacity) depend on both
quantum parameters $Q_b$ and $Q_c$.
Both expressions have the correct classical limits.

\begin{figure}[htb]
\includegraphics[width=0.5\linewidth]{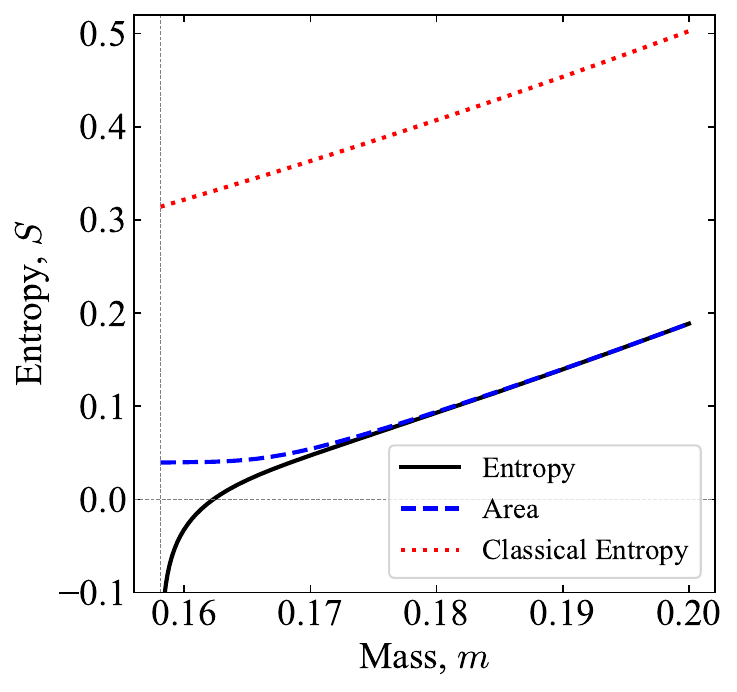}
\caption{\label{fig:entropy}Entropy and area versus mass for $Q_b =
  0.1$ and $Q_c = 10^{-6}$. Also shown is the classical entropy.}
\end{figure}

We are unable to integrate the expression for the entropy
(\ref{eq:entropy}) analytically and will resort to numerical
integration. 
In the asymptotic limit $S\to A/4$ and this will serve as our starting
point of integration.
Figure~\ref{fig:entropy} shows the entropy and area versus mass.
Also shown is the classical value of $S = 4\pi m^2$.

\section{Summary\label{sec:summary}}

We have previously derived an effective quantum corrected Schwarzschild
black hole using a deformed Poisson algebra inspired by the general
uncertainty principle~\cite{Fragomeno:2024tlh}. 
The solution resolves the classical curvature singularity and has all
the correct asymptotic and classical limits.
In this work, we have studied black hole, wormhole, and remnant
geometries. 

The line element is non-$t$-$r$-symmetric and replaces the classical
singularity with a coordinate singularity.
The coordinate singularity is interpreted as timelike and null
infinities which cause the two spacetimes to be causally disconnected
at $r=0$. 
This is in contradistinction to the usual geometry that describes two
causally connected universes by a bounce into a future incarnation
of the universe or a bounce back into our own universe.
Some of the results presented here may be applicable to other diagonal  
non-$t$-$r$-symmetric line elements~\cite{Modesto:2009ve,
Ashtekar:2020ckv, Gambini:2020nsf}.   

The metric is not of the Bardeen~\cite{BardeenBH},
Roman-Bergmann~\cite{PhysRevD.28.1265}, Frolov~\cite{Frolov:2016pav}
or Hayward~\cite{PhysRevLett.96.031103}, etc.\ type. 
The aforementioned class of regular black holes has been shown to have
unstable inner horizons~\cite{DiFilippo:2022qkl}. 
Our metric has no inner horizon and thus is not in this class.

The expansion scalars for a congruence of null geodesics have the
asymptotic limit of the classical case.
The expansion scalars vanish at both $r=0$ and the future null
infinity, and there is no caustic of the congruences.
Since the expansion scalars turn around at small $r$, the GUP
spacetime generates an effective repulsive effect.
The focusing theorem is not obeyed due to the violation of the null
energy condition.

The metric violates all the classical energy conditions associated
with the stress-energy-momentum tensor near $r=0$.
However, all energy conditions are satisfied at $r=0$, $r\to\infty$,
and the horizon (for black holes).
This confirms the expectation that at least one of the energy
conditions must be violated for regular black holes.
The negative radial pressure results in the dominant repulsive
behavior that prevents the formation of the spacetime singularity at
the center.  

An apparent worm hole solution is possible.
Like most wormholes the NEC, WEC, SEC, and DEC are violated near the 
throat. 
However, radial geodesics are unable to penetrate the wormhole throat
which is at timelike/null infinity.
The wormhole is not traversable in principle, if not in reality.

The metric is symmetric in $\pm r$.
Two bounded pseudo-Riemannian patches distinguished by $r > 0$ and
$r < 0$ are joining along their boundary at $r=0$.
In Schwarzschild coordinates the two patches are not properly glued at
$r = 0$; the spatial three dimensions are glued properly, but not
the time dimension.
The passage of a particle from the sheet $r > 0$ to $r < 0$ would go
through $t = \infty$, which is not a well defined part of the manifold.
In Kruskal coordinates, the determinant of the metric vanishes at $r =
0$ and the manifold is singular psuedo-Riemannian.

A finite mass zero-temperature gravitational remnant is possible.
Although the heat capacity is positive when the temperature vanishes, 
it is not clear that this would ensure stability.
\begin{acknowledgments}
We acknowledge the support of the Natural Sciences and Engineering
Research Council of Canada (NSERC). 
Nous remercions le Conseil de recherches en sciences naturelles et en
g{\'e}nie du Canada (CRSNG) de son soutien.
\end{acknowledgments}
\appendix 
\section{Derivation of the GUP metric from the modified Poisson bracket}
\label{appC}

In this appendix, we sketch out the derivation of the line element
(\ref{eq:diagonal}) and (\ref{eq:functions}) starting from the modified Poisson
bracket (\ref{eq:Poisson}).
Given the classical equations of motion $(i = b, c)$

\begin{equation}
\dot{q}_i = \left\{ q_i, H \right\}_\mathrm{cl} \quad \textrm{and} \quad
\dot{p}_i = \left\{ p_i, H \right\}_{cl}\, ,
\end{equation}
and the classical Poisson brackets (\ref{eq:PoissonClass}), the classical
algebra is modified to the effective bracket:

\begin{equation}
\left\{ q_i, p_j \right\}_\mathrm{eff} = \left\{ q_i, p_j \right\}_\mathrm{cl}
\left[ 1 + F(q,p,\beta_i) \right] \delta_{ij}\, ,
\end{equation}
where $F(q,p,\beta_i)$ is a function depending on the GUP modification to the
Poisson bracket. 
The new equations of motion are:
\begin{eqnarray}
\dot{q}_i & = & \left\{ q_i, H \right\}_\mathrm{eff} = \left\{q_i, H
\right\}_\mathrm{cl} \left[ 1 + F(q,p,\beta_i) \right]\, ,\\
\dot{p}_i & = & \left\{ p_i, H \right\}_\mathrm{eff} = \left\{p_i, H
\right\}_\mathrm{cl} \left[ 1 + F(q,p,\beta_i) \right]\, .
\end{eqnarray}
From the Poisson bracket (\ref{eq:Poisson}), the choice of
lapse~\cite{Fragomeno:2024tlh}   
\begin{equation}
\tilde{N} = \frac{\gamma\sqrt{p_c}}{\tilde{b}}\, ,
\end{equation}
and Hamiltonian~\cite{Fragomeno:2024tlh}
\begin{equation}
\tilde{H} = -\frac{1}{2\gamma} \left[ (\tilde{b}^2 + \gamma^2)
\frac{\tilde{p}_b}{\tilde{b}} + 2\tilde{c} \tilde{p}_c\right]\nonumber\, ,
\end{equation}
we obtain the following equations of motion in the interior:

\begin{eqnarray}
  \dot{b} & = & -\frac{b^2+\gamma^2}{2b} \left( 1 + \frac{\beta_b
    L_0^4}{p_b^2} b^2 \right)\nonumber\, ,\\ 
  \dot{p}_b & = & \frac{p_b(b^2-\gamma^2)}{2b^2} \left( 1 + \frac{\beta_b
    L_0^4}{p_b^2} b^2 \right)\nonumber\, ,\\ 
  \dot{c} & = & -2c \left( 1 + \frac{\beta_c
    L_0^4}{p_c^2} c^2 \right)\nonumber\, ,\\ 
  \dot{p}_c & = & 2p_c \left( 1 + \frac{\beta_c
    L_0^4}{p_c^2} c^2 \right)\, ,
\end{eqnarray}
where a dot denotes a derivative with respect to $\tilde{t}$.
Solving the equations of motion and fixing the integration constants by
matching to the classical limits gives

\begin{eqnarray}
b   & = & \gamma \sqrt{ \frac{2m}{\tilde{G}} - 1 }\nonumber\, ,\\ 
p_b & = & L_0 \tilde{G} \sqrt{\frac{2m}{\tilde{G}}-1}\nonumber\, ,\\
c   & = & -\frac{\gamma L_0 m}{\tilde{H}}\nonumber\, ,\\
p_c & = & \tilde{H}\, ,
\label{eq:eom}
\end{eqnarray}

\noindent
where $\tilde{G} = \sqrt{\tilde{t}^2+Q_b}$ and $\tilde{H} = \left( \tilde{t}^8 
+ Q_c m^2\right)^{1/4}$, with $Q_b$ and $Q_c$ defined in (\ref{eq:def}). 
Substituting the equations of motion (\ref{eq:eom}) in to the Kantowsk-Sachs
line element (\ref{eq:metric}) gives the metric for the interior of the GUP
black hole. 
The full GUP spacetime is derived by analytical extending the interior solution
using the transformations $\tilde{t} \to r$, $\tilde{r} \to t$, $\tilde{\theta}
\to \theta$, and $\tilde{\phi} \to \phi$, where $(t,r,\theta,\phi)$ are the
usual Schwarzschild coordinates. 
This gives the GUP spacetime line element (\ref{eq:diagonal}) and
(\ref{eq:functions}). 

\section{Painlev{\'e}-Gullstrand coordinates\label{appA}}

In this Appendix, we derive the Painlev{\'e}-Gullstrand line element
for a non-$t$-$r$-symmetric metric.
Following~\cite{Francis:2003rj}, first the most general nondiagonal
static spherically symmetric metric is developed.
Starting from the diagonal metric (\ref{eq:diagonal}), we apply the
transformation $u = \alpha t + B(r)$.
Without loss of generality, we take $\alpha = 1$ to obtain the
following general metric 

\begin{equation}
ds^2 = g_{00} du^2 + 2 B(r) du dr + \left( g_{11} +
\frac{B(r)^2}{g_{00}} \right) dr^2\, ,
\label{eq:general}
\end{equation}
where $u$ is a new time coordinate.
The Schwarzschild coordinates are recovered when $B(r) = 0$.

The Lagrangian for a free particle (massive or massless) can be
written as

\begin{equation}
\kappa = 2\mathcal{L}(x,\dot{x},\lambda)  = -g_{\mu\nu}  \dot{x}^\mu
  \dot{x}^\nu\,
= g_{00} \dot{u}^2 + 2B(r)\dot{u}\dot{r} + \left( g_{11} +
\frac{B(r)^2}{g_{00}} \right) \dot{r}^2\, ,
\end{equation}
where dots denote derivative with respect to the affine parameter
$\lambda$.
The Lagrangian is a constant of the motion which we denote by
$\kappa$; $\kappa = 1$ for massive particles (timelike)
and $\kappa = 0$ for massless particles (null).
The affine parameter can be used as the proper time for the timelike
case. 

In addition, $\partial\mathcal{L}/\partial\dot{u}$ is a constant
$E$ related to the energy:

\begin{equation}
\frac{\partial\mathcal{L}}{\partial\dot{u}} = g_{00} \dot{u} + B(r)
\dot{r} = E\, .
\label{eq:energy}
\end{equation}
For timelike motion $E$ is the energy per unit mass and for the null
motion $E$ is the energy, or frequency~\cite{Francis:2003rj}.

Substituting (\ref{eq:energy}) for $\dot{u}$ into the Lagrangian
gives $\dot{r}$ which happens to be independent of $B(r)$.
Then the result can be substituted back into (\ref{eq:energy}) to obtain
$\dot{u}$  in terms of $B(r)$.
The velocities with respect to the affine parameter are

\begin{eqnarray}
\dot{r} & = & \left( E^2 + \kappa \frac{g_{00}}{-g_{00} g_{11}}
\right)^{1/2}\, ,\\ 
\dot{u} & = & \frac{E}{g_{00}} \pm \frac{B(r)}{g_{00}} \left(
\frac{E^2 +\kappa g_{00}}{-g_{00} g_{11}} \right)^{1/2}\, .
\label{eq:u}
\end{eqnarray}
We next pick the coordinates system such that

\begin{equation}
\dot{u} = \frac{du}{d\lambda} = 1\, .
\end{equation}
Using this expression in (\ref{eq:u}) and solving for $B(r)$ we obtain

\begin{equation}
B(r) = \pm \left( \frac{-g_{00} g_{11}}{1+g_{00}} \right)^{1/2}
\left( 1 + g_{00} \sqrt{\frac{p}{\kappa}} \right)\, , 
\end{equation}
where $p=\kappa/E^2$.
For massive particles $\kappa=1$ and for the Painlv{\'e}-Gullstrand
metric $p=1$:

\begin{equation}
B(r) = \pm \sqrt{ (-g_{00} g_{11}) (1+g_{00}) } .
\end{equation}
We then substitute $B(r)$ into the general metric (\ref{eq:general}) to
obtain 

\begin{equation}
ds^2 = g_{00} d\tau^2 \pm 2\sqrt{(-g_{00} g_{11})(1+g_{00})} d\tau dr
- g_{00} g_{11} dr^2\, ,
\end{equation}
the Painlev{\'e}-Gullstrand metric for non-$t$-$r$-symmetric
diagonal metrics.
The advantage of this line element is that the time coordinate is the 
proper time of a radially infalling massive particle moving on a
geodesic and the hypersurfaces $\tau =$ constant are all intrinsically
flat. 
The usual Painlev{\'e}-Gullstrand metric is obtained for the
$t$-$r$-symmetric case of $g_{00} = -1/g_{11}$.

\section{Causal structure and Carter-Penrose diagram\label{appB}}

The global causal structure of the spacetime is studied by drawing
a Carter-Penrose diagram~\cite{CarterPenrose}.
In this Appendix we develop the coordinate transformations to enable
drawing such diagrams.
We will draw the diagrams using compactified null Kurskal coordinates. 
Starting from Schwarzschild coordinates $(t,r,\theta,\phi)$ the line
element is

\begin{equation}
ds^2 = g_{00} dt^2 + g_{11} dr^2 + g_{22} d\Omega^2\, ,
\end{equation}
where $-\infty < t < \infty$ and $r \ge0$.
The tortoise coordinate $r_*$ is given by integrating

\begin{equation}
\frac{dr_*}{dr} = \sqrt{\frac{g_{11}}{-g_{00}}}\, ,
\end{equation}
where $-\infty < r_* < \infty$ for $r > r_h$.
While $-\infty < r_* < 0$ in the Schwarzschild interior, we will
find $-\infty < r_* < \infty$ for $0 < r < r_h$ for our metric interior.
The two spacetime patches will be studied separately.
The metric in the tortoise coordinate becomes

\begin{equation}
ds^2 = g_{00} (dt^2 - dr_*^2) + g_{22} d\Omega^2\, ,
\end{equation}
where $g_{00}$ and $g_{22}$ are implicit functions of $r_*$ via $r = r(r_*)$.

Introducing the light-cone coordinates

\begin{equation}
u = t - r_* \quad\textrm{and}\quad v = t + r_*\, ,
\end{equation}
gives the metric

\begin{equation}
ds^2 = g_{00} du dv + g_{22} d\Omega^2\, ,
\end{equation}
where $-\infty < u < \infty$ and $-\infty < v < \infty$.
Now $g_{00}$ and $g_{22}$ are function of $u$ and $v$. 
These coordinates render the metric to be conformally flat in two-space.

The most general coordinate transformation which leaves this two-space
conformally flat is $V = V(v)$ and $U = U(u)$, where $U$ and $V$ are
arbitrary continuously differential functions.
The metric in the new coordinates becomes 

\begin{equation}
ds^2 = -F^2 dU dV + g_{22} d\Omega^2\, ,
\end{equation}
where

\begin{equation}
F^2 = - g_{00} \frac{\partial u}{\partial U} \frac{\partial
  v}{\partial V}
\end{equation}
and $F$ is implicitly define in terms of $U$ and $V$.

We define the null Kruskal coordinates

\begin{equation}
U = \mp e^{-u/k}\quad\textrm{and}\quad V = e^{v/k}\, ,
\end{equation}
where $k$ is a dimensional parameter that will be chosen to eliminate the
coordinate singularity at the horizon~\cite{Modesto:2008im}.
The upper sign is for the exterior and the lower sign for the interior
of the black hole.
Now

\begin{equation}
F^2 = k^2 g_{00} e^{-2 r_*/k}\, .
\end{equation}
The Kruskal coordinates $(t^\prime,r^\prime,\theta,\phi)$ are given
by $t^\prime = (U+V)/2$ and $r^\prime = (U-V)/2$ such that
$(t^\prime)^2 - (r^\prime)^2 = UV$.
In these coordinates, the metric assumes the conformally flat form
$ds^2 = F^2 \left[ -(dt^\prime)^2 + (dr^\prime)^2 \right]$.

The compactified null Kruskal coordinates are

\begin{equation}
\tilde{U} = \arctan \left( U\right)
    \quad\textrm{and}\quad \tilde{V} = \arctan\left(
    V\right)\, , 
\end{equation}
where

\begin{equation}
-\frac{\pi}{2} < \tilde{V} < \frac{\pi}{2}\, , \quad
-\frac{\pi}{2} < \tilde{U} < \frac{\pi}{2}\, , \quad
-\pi < \tilde{V} + \tilde{U} < \pi\, .
\end{equation}

Finally, we introduce the coordinates

\begin{equation}
  T = \frac{1}{2} (\tilde{V} + \tilde{U}) \quad\textrm{and}\quad
  R = \frac{1}{2} (\tilde{V} - \tilde{U})\, .
\end{equation}
Each point on the diagram, including the $r=0$ line corresponding to
the origin,  represents a finite radius two-sphere.
The metric depends explicitly on $g_{00}$ and implicitly through
coordinate transformations on the combination $\sqrt{-g_{11}/g_{00}}$. 

Now let's consider the tortoise coordinate for our metric.
The tortoise coordinate is given by integrating

\begin{equation}
  \frac{dr_*}{dr} =
  \left( 1 + \frac{Q_c m^2}{r^8} \right)^{1/4}
\frac{r}{\sqrt{r^2+Q_b} - 2m}\, .
\end{equation}
We are unable to integrate the tortoise equation analytically over the
entire range in $r$.
We integrate it numerically as described below.
To gain insight and assist the numerical integration, we consider some
analytical approximations.
Near the event horizon, we can neglect the $Q_c$ term and perform the
integration analytically to obtain

\begin{equation}
r_* = \sqrt{r^2+Q_b} + 2m \ln \left| \frac{\sqrt{r^2+Q_b}}{2m} -
1\right|\, .
\end{equation}
Choosing $k = 4m$, the metric function becomes

\begin{equation}
F^2 = \frac{32 m^3}{r} \left( 1 + \frac{Q_b}{r^2} \right)^{1/2}
\left( 1 + \frac{Q_c m^2}{r^8} \right)^{-1/4} \exp \left(
\frac{-\sqrt{r^2+Q_b}}{2m} \right)\, .
\label{eq:tortoise}
\end{equation}
The metric is now regular across the horizon.
In the classical limit, the spacetime is the same as the Schwarzschild
exterior.
The quantum corrections change the location of the horizon radius
and include a negligible $Q_c$-dependent factor in the exterior.

Now consider the interior.
Near the horizon the above approximation (\ref{eq:tortoise}) still holds.
However, near $r=0$ the $Q_c$ term dominates and we approximate

\begin{equation}
  \frac{dr_*}{dr} =
  \frac{(Q_c m^2)^{1/4}}{r}
  \frac{1}{\sqrt{r^2 + Q_b} - 2m}\, .
\end{equation}
Integration gives

\begin{equation}
r_* = \left( \frac{Q_c m^2}{r_h^8} \right)^{1/4} \left[ 2m \ln \left(
\sqrt{1 + \frac{Q_b}{r^2}}\right) + \frac{\sqrt{Q_b}}{2} \ln \left(
\frac{\sqrt{r^2 + Q_b} + \sqrt{Q_b}}{\sqrt{r^2 + Q_b} - \sqrt{Q_b}}
\right)  \right]\, .
\end{equation}
We see that both logarithms diverge as $r\to 0$ and it not possible to
pick a $k$ value that will eliminate both terms.
Consider the choice $k = 4 m (Q_c m^2/r_h^8)^{1/4}$, which gives

\begin{equation}
F^2 = \frac{16 m^2}{r} \left( \frac{Q_c m^2}{r_h^8} \right)^{1/2}
\left( 1 + \frac{Q_c m^2}{r^8} \right)^{-1/4} \left( \sqrt{r^2 + Q_b}
- 2m \right) \left( \frac{\sqrt{r^2 + Q_b}
  - \sqrt{Q_b}}{\sqrt{r^2 + Q_b} + \sqrt{Q_b}}
  \right)^{\frac{\sqrt{Q_b}}{4 m}}\, .
\end{equation}
In these coordinates, as $r\to 0$ the metric become purely a two-sphere:
$ds^2 = (Q_c m^2)^{1/4} d\Omega^2$.
Indeed, for all positive $k$ this is the case.

\begin{figure}[htb]
\includegraphics[width=0.5\linewidth]{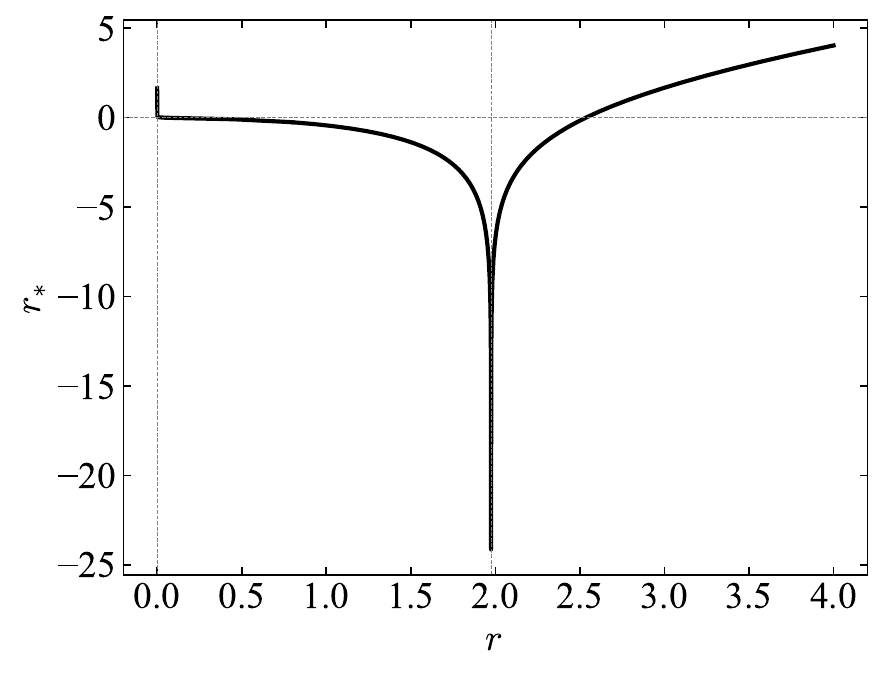}
\caption{\label{fig:tortoise}Tortoise coordinate $r_*$ versus $r$.
Discontinuities appear at $r = r_h$ and $r = 0$.}
\end{figure}

We now proceed with the numerical integration to obtain the tortoise
coordinate. 
In the exterior patch, the tortoise relation is integrated backwards
from $r = r_* = 350$ (representing infinity) to near $r_h$.
The numerical result for $r > r_h$ as shown in Fig.~\ref{fig:tortoise}
is visibly identical to the analytic approximation
(\ref{eq:tortoise}). 
In the interior patch, there is no natural point at which we know a
finite $r_*$ to start the integration.
We use the exterior approximation to obtain a point $r_*(r_0)$ in the
interior to start the integration.
We integrate forward to $r_h$ and obtain a result visibly identical to
the analytic approximation (\ref{eq:tortoise}).
We also integrate backwards to near $r=0$ for which we see a strong
positive discontinuity in $r_*$ at $r = 0$.
Figure~\ref{fig:tortoise} shows $r_*$ versus $r$.
Discontinuities appear at $r = r_h$ and $r = 0$.

\clearpage
\bibliographystyle{apsrev4-2}
\bibliography{gingrich}
\end{document}